\documentclass[twocolumn,showpacs,preprintnumbers,amsmath,amssymb,prl,
superscriptaddress]{revtex4}
\usepackage{graphicx}
\usepackage{dcolumn}
\usepackage{bm}
\usepackage{amsmath}
\usepackage{hyperref}

\def\rnum#1{\expandafter{%
\romannumeral #1}}
\def\Rnum#1{\uppercase\expandafter{%
\romannumeral #1}}

\newcommand{\bol}[1]{\boldsymbol #1}

\begin{document}

\title{
Theory of Raman Scattering in One-Dimensional Quantum 
Spin-$\frac{\bol 1}{\bol 2}$ Antiferromagnets
}

\author{Masahiro Sato}
\affiliation{Department of Physics and Mathematics, Aoyama Gakuin University, 
Sagamihara, Kanagawa 229-8558, Japan}
\author{Hosho Katsura}
\affiliation{Department of Physics, Gakushuin University, Mejiro, Toyoshima-ku, Tokyo 171-8588, Japan}
\author{Naoto Nagaosa}
\affiliation{Department of Applied Physics, University of Tokyo, 
Hongo, Bunkyo-ku, Tokyo 113-8656, Japan}
\affiliation{CMRG and CERG, RIKEN-ASI, Wako, Saitama 351-0198, Japan}
\date{\today}

\begin{abstract}
We study theoretically the Raman scattering spectra in the 
one-dimensional (1D) quantum spin-$\frac{1}{2}$ antiferromagnets. 
The analysis reveals that their low-energy 
dynamics is exquisitely sensitive to various perturbations to 
the Heisenberg chain with nearest-neighbor exchange interactions, 
such as magnetic anisotropy, longer-range exchange
interactions, and bond dimerization. 
These weak interactions are mainly responsible for 
the Raman scattering and give rise to different types of spectra 
as functions of frequency, temperature, and external field. 
In contrast to the Raman spectra in higher dimensions in which 
the two-magnon process is dominant, 
those in 1D antiferromagnets 
provide much richer information on these perturbations. 
\end{abstract}




\maketitle

\textit{Introduction}: 
Quantum 
antiferromagnets have long attracted much attention 
as a laboratory to study quantum many-body effects. 
Experimentally, several 
techniques are available to investigate 
them; measurements of magnetic susceptibility, specific heat, 
and spectra of neutron scattering, NMR, and ESR. 
Recently, the optical spectra~\cite{Devereaux,Lemmens,Katsura,Wang} 
have also turned out to be a powerful tool to study quantum spin dynamics. 
An example is electromagnon spectroscopy of multiferroics, 
where the one-magnon process
is activated by an electric field in infrared 
absorption due to the magnetostriction mechanism~\cite{Wang}. 
Raman scattering~\cite{Devereaux,Lemmens,Fleury,Moriya}, 
on the other hand, is usually considered to detect two-magnon excitations 
in antiferromagnetic (AF) ordered phases, which has been utilized to estimate 
the strength of the exchange interaction.

In one-dimensional (1D) systems where quantum fluctuations are 
much enhanced, such a simple magnon picture fails miserably. 
A canonical model for 1D quantum antiferromagnets is 
the spin-$\frac{1}{2}$ Heisenberg Hamiltonian, 
\begin{eqnarray}
\label{eq:Heisenberg}
{\cal H}_0 &=& J \sum_j {\bol S}_j \cdot {\bol S}_{j+1},
\end{eqnarray}
where $J$ ($>0$) is the exchange interaction between neighboring spins. 
The low-energy physics of this system is described by a Tomonaga-Luttinger 
(TL) liquid with gapless spinon excitations~\cite{Giamarchi} 
instead of magnons. In real systems, however, additional small 
perturbations ${\cal V}$ always exist, e.g., spin-orbit interaction, 
magnetic anisotropy, disorder, longer-range exchange interactions, and 
also spin-lattice coupling leading to the bond dimerization 
(spin-Peierls instability). Despite the smallness of 
these interactions, they are crucial 
for the quantum dynamics of the system, and are the subject of 
intensive studies.


Unfortunately, experimental
signatures of these small perturbations ${\cal V}$ are often difficult to study
because conventional experiments probe quantities that are dominated
by the Heisenberg term~(\ref{eq:Heisenberg}) of the Hamiltonian. 
Therefore, it is highly desirable to have experimental probes 
that reveal the physical processes associated with the small perturbations 
${\cal V}$ in 1D antiferromagnetic systems. 
In this Letter, we show that Raman
scattering from 1D spin-$\frac{1}{2}$ antiferromagnets provides 
such an experimental probe.

It has been considered so far that the Raman scattering in 1D magnets 
is not so useful compared to other conventional methods 
although some of the experimental and theoretical works 
exist~\cite{Lemmens,Muthukumar,Lemmens2,Loosdrecht,Ruckamp,
Schmidt,Schmidt2,Orignac}. 
This is because the Hamiltonian ${\cal H}_0$ and 
the corresponding Raman operator ${\cal R}_{0} \propto {\cal H}_0$ 
[see Eq.~(\ref{eq:Raman})] commute with each other, 
and hence no Raman scattering occurs without additional interactions.  
Furthermore, these perturbations ${\cal V}$, which determine 
the Raman scattering spectra (RSS), remain rather uncertain in most cases. 
However, this does not necessarily mean that the RSS is useless 
in these systems. In fact, once theoretical predictions on the RSS for each 
interaction ${\cal V}$ are available, 
RSS can provide useful information on ${\cal V}$ as we will see later.

The results of our analysis based on field-theory and nonperturbative methods 
are summarized in Tables \ref{tab:table1} and 
Fig.~\ref{fig:RSS_Gapless} for gapless cases, 
and Table~\ref{tab:table2} and Fig.~\ref{fig:Gapful} for gapped cases, 
respectively. Comparing these predictions with the 
observed temperature, frequency, and 
magnetic-field dependence of the RSS, one can obtain 
detailed information on ${\cal V}$. 
The results will be explained below. 

\textit{Definition of RSS}:
Let us start from the definition of the RSS and the Raman operator. 
The RSS 
is proportional to the dynamical structure 
factor of the Raman operator ${\cal R}$, namely,
\begin{eqnarray}
\label{eq:Intensity}
I(\omega) &=& \frac{1}{2\pi}\int^\infty_{-\infty} dt \,\,e^{i\omega t}
\langle {\cal R}(t){\cal R}(0)\rangle,
\end{eqnarray}
where $\omega=\omega_i-\omega_s$ and $\omega_{i(s)}$ is the energy of incident 
(scattered) photon. In Mott-insulating systems, 
the Raman operator~\cite{Devereaux,Lemmens,Fleury,Moriya} usually 
has the form 
\begin{eqnarray}
\label{eq:Raman}
{\cal R} &=& \sum_{{\bol r}_1,{\bol r}_2}({\bol e}_i\cdot {\bol r}_{12})
({\bol e}_s\cdot {\bol r}_{12}) A({\bol r}_{12}) 
{\bol S}_{{\bol r}_1} \cdot {\bol S}_{{\bol r}_2},
\end{eqnarray}
where ${\bol e}_{i(s)}$ is the polarization direction of the incident 
(scattered) photon and ${\bol r}_{12}={\bol r}_1-{\bol r}_2$. 
Therefore, the RSS strongly depends on the direction of 
applied and observed electromagnetic waves and the crystal structure of 
magnets. The factor $A({\bol r}_{12})$ is difficult to accurately determine, 
but the ratio between the factors on different bonds is known to be 
of the same order as that between the exchange couplings on those bonds. 
From Eqs.~(\ref{eq:Intensity}) and (\ref{eq:Raman}), 
one can easily find that the intensity $I(\omega)$ is 
unchanged when ${\cal R}$ is replaced with the modified Raman operator
\begin{eqnarray}
\label{eq:Raman2}
{\cal R}'&=&  {\cal R}-C{\cal H},
\end{eqnarray}
where $C$ is arbitrary real constant and ${\cal H}$ is the Hamiltonian of 
the target magnet. We can therefore 
adopt ${\cal R}'$ to make the calculation of $I(\omega)$ easier. 

\textit{Analysis}:
The low-energy physics 
of the Heisenberg chain~(\ref{eq:Heisenberg}) 
with/without an easy-plane anisotropy 
${\cal V}_1=-J\Delta\sum_j S_j^zS_{j+1}^z$ and a Zeeman term 
is well described by the TL-liquid theory~\cite{Giamarchi}. 
The low-energy effective Hamiltonian is 
identical to the free boson theory
\begin{eqnarray}
\label{eq:Effective}
{\cal H}_0^{\rm eff} &=& \int dx \frac{v}{2}\left[K^{-1}(\partial_x\phi)^2
+K(\partial_x\theta)^2\right],
\end{eqnarray}
where $x=ja_0$ ($a_0$ is lattice spacing), $v$ is the spinon velocity, 
$K$ is the TL-liquid parameter, and 
$(\phi,\theta)$ 
is the canonical pair of bosonic fields. 
The parameter $K=1$ at the SU(2) symmetric point, 
while an anisotropy ${\cal V}_1$ or an external field $H$ 
usually increases the value of $K$, i.e., $K>1$. Note that 
${\cal V}_1$ is the perturbation in the sense that it violates 
the commuting property of the Hamiltonian ${\cal H}_0+{\cal V}_1$ and 
the Raman operator ${\cal R}$. 
Spin and dimer operators can also be bosonized as 
$S^\alpha_j\approx {\cal J}^\alpha(x)+(-1)^j {\cal N}^\alpha(x)$ and 
$(-1)^j {\bol S}_j \cdot {\bol S}_{j+1}\approx d\sin(\sqrt{2\pi}\phi)+\cdots$, 
where ${\cal N}^+=b_0 e^{i\sqrt{2\pi}\theta}+\cdots$, 
${\cal J}^+=b_1 e^{i\sqrt{2\pi}\theta}\cos(\sqrt{2\pi}\phi)+\cdots$, 
${\cal N}^z=a_1\cos(\sqrt{2\pi}\phi)+\cdots$, and 
${\cal J}^z=a_0\partial_x\phi/\sqrt{2\pi}$. 
The anisotropy and field dependence of parameters 
$a_1$, $b_{0,1}$, $d$, $K$ and $v$ is accurately 
evaluated~\cite{Lukyanov,H-F,T-S}. 
The bosonization approach therefore enables us to 
estimate the effects of several perturbations on the RSS $I(\omega)$ 
with reasonable accuracy in the low-energy region, i.e., $T,|\omega|\ll J$.

\begin{figure}
\begin{center}
\includegraphics[width=8.6cm]{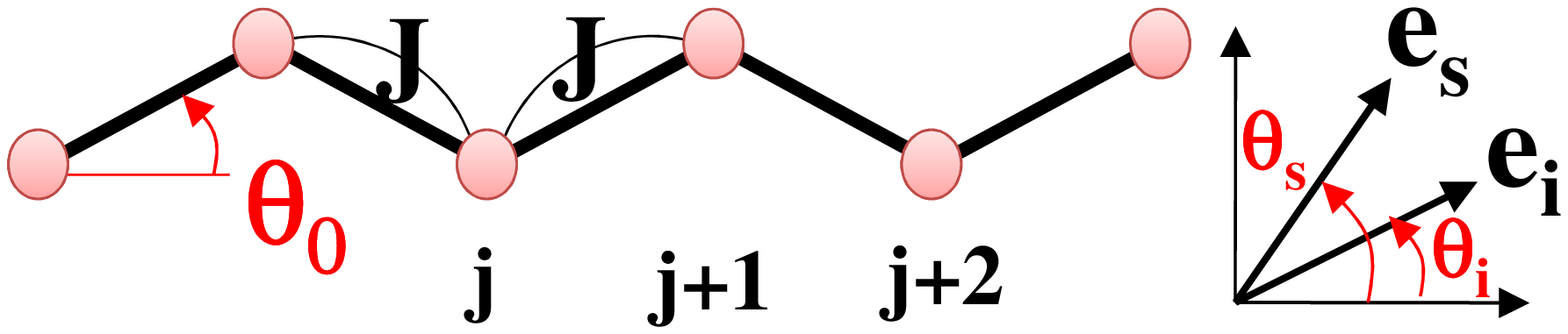}
\end{center}
\caption{(color online) 1D antiferromagnet with tilting angle $\theta_0$ 
and polarization directions of incident and scattering photon ${\bol e}_{i,s}$ 
with angles $\theta_i$ and $\theta_s$.}
\label{fig:bond_tilting}
\end{figure}

\begin{figure}
\begin{center}
\includegraphics[width=8.6cm]{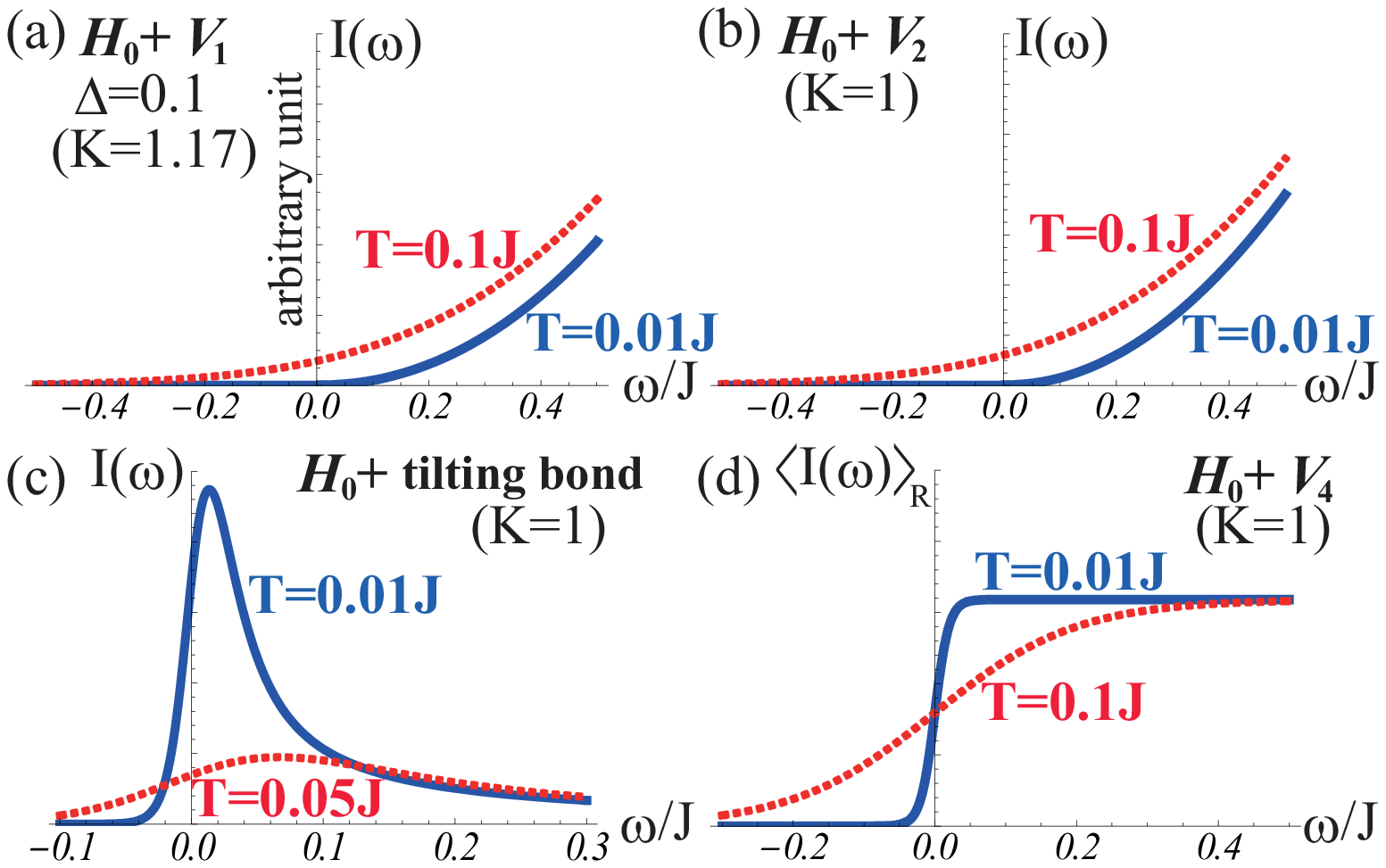}
\end{center}
\caption{(color online) RSS $I(\omega)$ for 1D Heisenberg 
magnet~(\ref{eq:Heisenberg}) with additional perturbation ${\cal V}$s. 
Panels (a), (b), (c) and (d) correspond to 
the cases of ${\cal V}_1$, ${\cal V}_2$, the bond tilting in 
Fig.~\ref{fig:bond_tilting}, and ${\cal V}_4$, respectively. 
All the continuous spectra come from multiple spinon states 
in the TL liquid phase. 
}
\label{fig:RSS_Gapless}
\end{figure}

\begin{figure}
\begin{center}
\includegraphics[width=8.6cm]{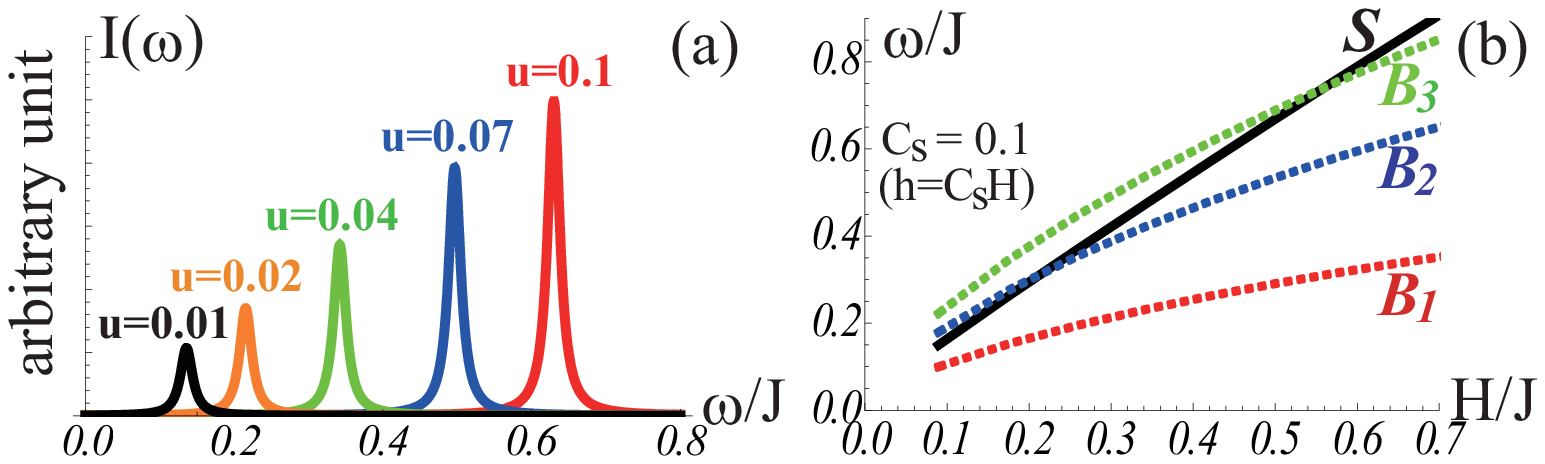}
\end{center}
\caption{(color online) (a) $u$ dependence of the singlet-breather 
(${\cal B}_2$) peak in $I(\omega)$ of 
the dimerized magnet ${\cal H}_0+{\cal V}_5$ at $T=0$, 
where $\delta$ functions are broadened. The peak position $\omega=E_2$ 
and its weight are both proportional to $u^{2/3}$. 
This peak is stable against an applied field $H$. 
The weight of the continuum spectrum with $\omega\geq 2E_s$ is much 
smaller than that of the ${\cal B}_2$ peak, and the former is omitted here. 
(b) $H$ dependence of peak positions of 
each particle in $I(\omega)$ of the magnet ${\cal H}_0+{\cal V}_6$ at $T=0$. 
Some level crossings between the soliton ${\cal S}$ and 
breather ${\cal B}_n$ peaks occur with increasing $H$. 
The continuum spectra are omitted.}
\label{fig:Gapful}
\end{figure}

\textit{Gapless Cases}: 
Let us study four realistic perturbations ${\cal V}$ that do not 
violate the TL-liquid phase; XXZ anisotropy ${\cal V}_1$, 
longer-range exchange couplings 
${\cal V}_2=\sum_{n\geq 2}\sum_{j}J_n {\bol S}_j\cdot{\bol S}_{j+n}$ 
($|J_n|\ll J$), a bond tilting in Fig.~\ref{fig:bond_tilting}, 
and a random bond alternation 
${\cal V}_4=\sum_jJ (-1)^j u_j {\bol S}_j\cdot{\bol S}_{j+1}$ with 
$u_j$ being the randomly distributed lattice distortion ($|u_j|\ll 1$). 
The results are summarized in 
Table~\ref{tab:table1} and Fig.~\ref{fig:RSS_Gapless}.

\begin{table*}
\caption{\label{tab:table1}Properties of RSS $I(\omega)$ in gapless cases of 
1D Heisenberg magnet ${\cal H}_0$ with perturbation ${\cal V}$. 
Constants $c_{1-12}$ depends on 
the TL-liquid parameter $K$, 
the spinon velocity $v$, the magnetization $M=\langle S_j^z\rangle$, 
the lattice spacing $a_0$, etc. The value of $K$ is unity 
at the SU(2)-symmetric case, while an easy-plane XXZ anisotropy 
or a magnetic field $H$ increases it, i.e., $K>1$. 
}
\begin{ruledtabular}
\begin{tabular}{llllll}
perturbation ${\cal V}$  &  bosonized form of ${\cal V}$ & scaling dimension  
& RSS $I(\omega)$ & main effect of field $H$\\ 
\hline
\begin{tabular}{ll}
XXZ anisotropy \\ 
${\cal V}_1=-J\Delta\sum_{j} S_j^z S_{j+1}^z$\\
\end{tabular}    
& 
$J\Delta({\cal N}^z{\cal N}^z-{\cal J}^z{\cal J}^z)$
& 
\begin{tabular}{ll}
$2K$ (${\cal N}^z$ term) \\
$2$ (${\cal J}^z$ and ${\cal N}^z$ terms)\\
\end{tabular}
& 
\begin{tabular}{ll}
$c_1\omega^{4K-2}+c_2\omega^2$ ($\beta\omega\gg 1$)\\
$c_3T^{4K-2}+c_4T^2$ ($\beta\omega\ll 1$)\\
\end{tabular}
&
\begin{tabular}{ll}
$c_{1,3}$ terms disappear \\
in $\omega\alt\omega_{1}=4\pi Mv/a_0$ \\
\end{tabular}
\\
\hline
\begin{tabular}{ll}
longer-range coupling \\  
${\cal V}_2=\sum_{j}J_n {\bol S}_j\cdot{\bol S}_{j+n}(n\geq2)$\\
\end{tabular}
& 
\begin{tabular}{ll}
$c_{xy}({\cal J}_R^+{\cal J}_L^-+{\rm h.c.})/2$ \\
$+c_z{\cal J}_R^z{\cal J}_L^z$\\
\end{tabular}
& 
\begin{tabular}{ll}
$2K$ ($c_{xy}$ term) \\
$2$ ($c_z$ term)  \\
\end{tabular}
& 
\begin{tabular}{ll}
$c_5\omega^{4K-2}+c_6\omega^2$ ($\beta\omega\gg 1$) \\ 
$c_7T^{4K-2}+c_8T^2$ ($\beta\omega\ll 1$) \\
\end{tabular}
&
\begin{tabular}{ll}
$c_{5,7}$ terms disappear \\
in $\omega\alt\omega_{1}$\\
\end{tabular}
\\
\hline
tilting bond in Fig.~\ref{fig:bond_tilting} 
& 
\begin{tabular}{ll}
$\sin(2\theta_0)\sin(\theta_i+\theta_s)$\\ 
$\times d\sin(\sqrt{2\pi}\phi)$ \\
\end{tabular}
& 
$K/2$
& 
\begin{tabular}{ll}
$c_9\omega^{K-2}$ ($\beta\omega\gg 1$) \\
$c_{10}T^{K-2}$ ($\beta\omega\ll 1$)\\
\end{tabular}
& 
\begin{tabular}{ll}
$c_{9,10}$ terms disappear \\
in $\omega\alt\omega_{2}=2\pi Mv/a_0$   \\
\end{tabular}
\\
\hline
\begin{tabular}{ll}
random dimerization   \\
${\cal V}_4=\sum_j J (-1)^j u_j {\bol S}_j\cdot{\bol S}_{j+1}$\\
\end{tabular}
& 
$Ju_jd\sin(\sqrt{2\pi}\phi)$
&
$K/2$ 
& 
\begin{tabular}{ll}
$c_{11}\omega^{K-1}$ ($\beta\omega\gg 1$) \\
$c_{12}T^{K-1}$ ($\beta\omega\ll 1$) \\
\end{tabular}
&  
value of $K$ increases
\end{tabular}
\end{ruledtabular}
\end{table*}

Generally ${\cal V}_{1,2}$ are always present in real compounds. 
The main part of the bosonized ${\cal V}_2$ is 
\begin{eqnarray}
\label{eq:long-range}
\int \frac{dx}{a_0}\,\, \frac{c_{xy}}{2}({\cal J}_R^+{\cal J}_L^-+{\rm h.c.})
+c_{z}{\cal J}_R^z{\cal J}_L^z
\end{eqnarray}
where ${\cal J}_{R(L)}^\alpha$ is the right (left) moving part of 
${\cal J}^\alpha$, and constants $c_{xy}=c_z$ depend on $J$ and $J_n$. 
Similarly, we obtain 
${\cal V}_1\approx-J\Delta\int \frac{dx}{a_0}({\cal J}^z{\cal J}^z
-{\cal N}^z{\cal N}^z)$. 
From Eq.~(\ref{eq:Raman2}), we can make ${\cal R}'$ proportional to 
${\cal V}_{1,2}$ 
if ${\bol e}_{i,s}$ are set parallel to 
${\bol r}_j-{\bol r}_{j+1}$. Applying the standard technique based on 
the bosonization and conformal field theory~\cite{Giamarchi}, 
we can calculate the Raman intensity of the Heisenberg chain 
${\cal H}_0$ with ${\cal V}_1$ or ${\cal V}_2$ for arbitrary frequency 
$\omega$ and temperature $T=1/\beta$. The result for the case of 
${\cal V}_2$ is 
\begin{eqnarray}
\label{eq:Raman_V1}
I(\omega) &\propto&  2c_{xy}^2 F(\omega,\beta,K)
+c_z^2 F(\omega,\beta,1+\epsilon)|_{\epsilon\to0}
\end{eqnarray}
where $F(x,y,z)=\frac{a_0}{2\pi v}(1-e^{-xy})^{-1}
\sin(2\pi z){\rm Im}[B(-i\frac{xy}{4\pi}+z,1-2z)^2]
(\frac{2\pi a_0}{y v})^{4z-2}$ and 
$B(x,y)$ is the Beta function. 
We also have $I(\omega)\propto \frac{1}{2}\pi^2 a_1^4 F(\omega,\beta,K)
+K^2(\frac{1}{2\pi}+\frac{\pi}{2}a_1^2)^2 F(\omega,\beta,1+\epsilon)$ 
for the case of ${\cal V}_1$. 
The intensities of these two cases are presented in 
Fig.~\ref{fig:RSS_Gapless} (a) and (b), and they show that 
$I(\omega)$ is a monotonically increasing function of $\omega$, 
and it remains finite in the limit $\omega/J\to 0$ at finite temperatures. 
These properties are at least qualitatively consistent with 
the experimental result of the paramagnetic phase of 
$\rm CuGeO_3$~\cite{Muthukumar,Lemmens2}, 
in which $J\sim 150$K and $J_2\sim 30$K. 
We note that ${\cal V}_1$ and ${\cal V}_2$ give the similar behavior 
in the RSS, but they can be distinguished by 
other physical quantities such as susceptibilities. 
When we apply a magnetic field $H$ and a magnetization 
$M=\langle S_j^z\rangle$ appears, 
$B(-i\frac{\beta\omega}{4\pi}+K,1-2K)^2$ in the first term 
$F(\omega,\beta,K)$ of $I(\omega)$ is changed into 
$B(-i\frac{\beta(\omega+4\pi Mv/a_0)}{4\pi}+K,1-2K)
B(-i\frac{\beta(\omega-4\pi Mv/a_0)}{4\pi}+K,1-2K)$ 
for both cases of ${\cal V}_{1,2}$. 
As a result, the RSS weight of this term 
becomes nearly zero in the low-frequency region 
$\omega\alt\omega_1=4\pi Mv/a_0$ at low temperatures $T\ll J$. 
For instance, in the case of ${\cal V}_2$, 
only the $c^2_z$ term survives in Eq.~(\ref{eq:Raman_V1}). 

We next consider 1D magnets with a tilting bond as in 
Fig.~\ref{fig:bond_tilting}. In fact, tilting structures with 
a small angle $\theta_0$ exist in 
several cuprate magnets such as Cu benzoate~\cite{Dender}, 
$\rm KCuGaF_6$~\cite{Umegaki}, and 
$\rm [PM]Cu(NO_3)_2(H_2O)_2$ (PM=pyrimidine)~\cite{Feyerherm}. 
In this system, the Hamiltonian is the same 
as Eq.~(\ref{eq:Heisenberg}) and hence a TL-liquid state survives. 
However, if we fix ${\bol e}_{i,s}$ as in Fig.~\ref{fig:bond_tilting}, 
the Raman operator becomes different from that of 
the case without a tilting bond. 
Tuning the value of $C$ in Eq.~(\ref{eq:Raman2}), we obtain
\begin{eqnarray}
\label{eq:Raman_V2}
{\cal R}' &\propto& \sin(2\theta_0)\sin(\theta_i+\theta_s)
\sum_j(-1)^j{\bol S}_j\cdot{\bol S}_{j+1}.
\end{eqnarray}
This is nothing but a dimerization operator and does not commute with 
${\cal H}_0$. Using this operator, we obtain 
$I(\omega)\propto \sin^2(2\theta_0)\sin^2(\theta_i+\theta_s)d^2 
F(\omega,\beta,K/2)$ that is depicted in 
Fig.~\ref{fig:RSS_Gapless} (c)~\cite{note1}. 
The RSS rapidly increases around $\omega=T$ at low temperatures, 
and the form is quite different from the case of ${\cal V}_{1,2}$. 
Physically the origin of this spectrum 
is two-spinon states. We emphasize that the strength of $I(\omega)$ 
can be controlled by tuning angles $\theta_{i,s}$. Similarly to the case of 
${\cal V}_{1,2}$, a magnetic field $H$ makes the weight of $I(\omega)$ 
absent in the region $\omega\alt\omega_2=2\pi Mv/a_0$.

A random dimerization is expected to be present in the 
higher-temperature paramagnetic phase of spin-Peierls compounds 
such as $\rm CuGeO_3$~\cite{Muthukumar} and $\rm TiOCr$~\cite{Ruckamp}. 
In this case, the Raman operator ${\cal R}'$ is proportional to 
${\cal V}_4/J\approx \int \frac{dx}{a_0}u_j d\sin(\sqrt{2\pi}\phi)$. 
Under the assumption that $u_j$ is a sufficiently small perturbation from 
${\cal H}_0$ and $\langle u_j u_k\rangle_R={\bar u}^2\delta_{jk}$ 
($\langle \cdots\rangle_R$ stands for the average over the randomness). 
the Raman intensity $\langle I(\omega) \rangle_R$ is reduced to 
a {\it local} correlator $\propto 
{\bar u}^2 d^2 \int \frac{dt}{2\pi} e^{i\omega t}
\langle \sin(\sqrt{2\pi}\phi(t))\sin(\sqrt{2\pi}\phi(0))\rangle$. 
It is calculated as 
\begin{eqnarray}
\label{eq:random_R}
\frac{{\bar u}^2d^2a_0}{4\pi v}
\left(\frac{2\pi a_0}{\beta v}\right)^{K-1}e^{\frac{\beta\omega}{2}}
B\left(\frac{K}{2}-i\frac{\beta\omega}{2\pi},
\frac{K}{2}+i\frac{\beta\omega}{2\pi}\right),
\end{eqnarray}
which is shown in Fig.~\ref{fig:RSS_Gapless} (d). 
The $\omega$ dependence of $\langle I(\omega)\rangle_R$ 
is negligible in $\omega>T$. 
Such a spectrum with a small slope is observed in the paramagnetic phase 
of $\rm CuGeO_3$ (see the region $\omega\alt 50{\rm cm}^{-1}$ in Fig.~1 
of Ref.~\onlinecite{Muthukumar}), and therefore the spectrum might contain 
the contribution from ${\cal V}_4$. 
An applied field $H$ does not affect the form of 
$\langle I(\omega)\rangle_R$ much, 
but it slightly varies parameters $(K,v,a_1,b_{0,1},d)$.

\textit{Gapped Cases}: Let us now discuss another kind of typical 
perturbations ${\cal V}$ that break the TL liquid in ${\cal H}_0$ and 
open a finite excitation gap: a static bond alternation (dimerization) 
term ${\cal V}_5=\sum_j J (-1)^ju {\bol S}_j\cdot{\bol S}_{j+1}$, 
and uniform and staggered Zeeman terms ${\cal V}_6=
-\sum_j HS_j^z+(-1)^j hS_j^x$ induced by an applied field $H$. 
The results are summarized in Table~\ref{tab:table2} and 
Fig.~\ref{fig:Gapful}.

\begin{table*}
\caption{\label{tab:table2}Properties of RSS $I(\omega)$ in gapped cases of 
1D Heisenberg magnet ${\cal H}_0$ with perturbation ${\cal V}$ at $T=0$. 
In the case of ${\cal V}_5$, the soliton mass, and first and second breather 
masses are respectively evaluated as $E_s\approx1.5 u^{2/3}J$, $E_1=E_s$ and 
$E_2=\sqrt{3}E_s$, while $E_s\approx 2.1(h/J)^{2/3}J$ 
in the case of ${\cal V}_6$ with a small field $H\ll J$. 
Note that in the case of ${\cal V}_5$, $E_2$ becomes larger than 
$\sqrt{3}E_s$~\cite{Schmidt} if the marginal operator, 
neglected in the SG model ${\cal H}_5$, is taken into account. 
On the other hand, a small next-nearest-neighbor AF coupling $J_2$ weakens 
the effect of the marginal term~\cite{Schmidt}.  
}
\begin{ruledtabular}
\begin{tabular}{lllll}
perturbation ${\cal V}$  &  bosonized form of ${\cal V}$ 
& Raman active mode (its main origin) 
 & peak positions for each mode & \\ 
\hline
\begin{tabular}{ll}
static dimerization \\
${\cal V}_5=\sum_jJ(-1)^ju {\bol S}_j\cdot{\bol S}_{j+1}$
\end{tabular}
&   
$Jud\sin(\sqrt{2\pi}\phi)$ 
& 
\begin{tabular}{ll}
second breather ($\sin(\sqrt{2\pi}\phi)$ term)\\
${\cal S}$-$\bar{\cal S}$ continuum ($\sin(\sqrt{2\pi}\phi)$ term)\\
\end{tabular}
& 
\begin{tabular}{ll}
$\omega=E_2$ (stable against $H$) \\
$\omega\geq 2 E_s$  
\end{tabular}
&
\\
\hline
\begin{tabular}{ll}
uniform and staggered\\
Zeeman terms\\
${\cal V}_6=-\sum_j HS_j^z+(-1)^j hS_j^x$\\
\end{tabular}
&  
\begin{tabular}{ll}
$-Ha_0\partial_x\phi/\sqrt{2\pi}$ \\
$-h b_0\cos(\sqrt{2\pi}\theta)$ \\
\end{tabular}
& 
\begin{tabular}{ll}
soliton, antisoliton (tilting bond)\\
odd-th breathers ($\partial_x\phi$ term)\\
even-th breathers ($\cos(\sqrt{2\pi}\theta)$ term)\\
\end{tabular}
& 
\begin{tabular}{ll}
$\omega=(E_s^2+(2\pi Mv/a_0)^2)^{1/2}$  \\
$\omega=E_{2n+1}$\\
$\omega=E_{2n}$ \\
\end{tabular}
&   
\\
\end{tabular}
\end{ruledtabular}
\end{table*}

In the case of ${\cal V}_5$, the effective Hamiltonian becomes 
an exactly solvable sine-Gordon (SG) model, 
\begin{eqnarray}
{\cal H}_5&=&{\cal H}_0^{\rm eff}+\int \frac{dx}{a_0} ud\sin(\sqrt{2\pi}\phi). 
\end{eqnarray}
There are three kinds of massive excitations: soliton (${\cal S}$), 
antisoliton ($\bar{\cal S}$), and some breathers (${\cal B}_n$) 
that are the soliton-antisoliton bound states. 
The mass of the soliton $E_s$ is equal to that of antisoliton, 
and it is given by~\cite{T-S}
\begin{eqnarray}
\label{eq:solitonmass}
\frac{E_s}{J}= \frac{v}{Ja_0}\frac{2}{\sqrt{\pi}}
\frac{\Gamma(\frac{K}{8-2K})}{\Gamma(\frac{2}{4-K})}
\left[\frac{Ja_0}{v}\frac{\pi d}{2}
\frac{\Gamma(\frac{4-K}{4})}{\Gamma(K/4)}\right]^{\frac{2}{4-K}},
\end{eqnarray}
where $\Gamma(x)$ is the Gamma function. The $n$-th breather's mass $E_n$ 
is related to $E_s$ via $E_n=2E_s\sin[n\pi/(8/K-2)]$ with 
$n=1,\cdots,[4/K-1]$. 
The SU(2)-symmetric dimerized chain with $K=1$ has 
only two breathers ${\cal B}_{1,2}$. 
Three particles ${\cal S}$, $\bar{\cal S}$ and ${\cal B}_1$ corresponds to 
massive spin-1 triplet excitations with $S^z=+1$, $-1$, and $0$, respectively, 
while ${\cal B}_2$ is regarded as a singlet excitation with $S=0$. The soliton 
mass is evaluated as $E_s\approx3.5 (du)^{2/3}J$ with 
$d\approx 0.3$~\cite{T-S} at the SU(2) point. 
From Eq.~(\ref{eq:Raman}), the RSS 
is proportional to the dynamical structure factor of $\sin(\sqrt{2\pi}\phi)$. 
To accurately evaluate such dynamical correlators of the SG model 
at the low-energy region, we utilize the form-factor approach~\cite{E-K,K-E} 
which is reliable when $T/J\ll 1$.   
From this approach, the lowest-frequency contribution of 
$I(\omega)$ is shown to be a $\delta$-functional peak of the singlet 
breather ${\cal B}_2$ at $\omega=E_2$, and the second lowest one is 
given by the continuum spectrum of soliton-antisoliton scattering states 
with $\omega\geq 2E_s$. The weight of each contribution 
can also be exactly calculated by the form-factor method. 
In particular, the weight of the singlet breather is proportional 
to $(E_sa_0/v)^{K}$, and is much larger than that of the continuum. 
The ${\cal B}_2$ peak and its weight in $I(\omega)$ are shown in 
Fig.~\ref{fig:Gapful} (a). The distortion ($u$) dependence of this peak 
can be compared to Raman scattering experiments for spin-Peierls magnets, 
$\rm CuGeO_3$~\cite{Muthukumar,Loosdrecht}, $\rm TiOCr$~\cite{Ruckamp}, etc. 
The ${\cal B}_2$ peak of $I(\omega)$ is stable against an applied field $H$ 
if $H$ is smaller than the critical field $H_c=E_s$.

A staggered magnetic field $h$ emerges as we apply a uniform field 
$H$ to magnets with a staggered gyromagnetic tensor~\cite{O-A}. 
Typical examples are Cu benzoate~\cite{Dender}, 
$\rm KCuGaF_6$~\cite{Umegaki}, 
and $\rm [PM]Cu(NO_3)_2(H_2O)_2$~\cite{Feyerherm}, 
in which a tilting structure is also present (as we discussed). 
In these compounds, $h\approx c_s H$ ($|c_s|\ll 1$) is realized. 
The bosonized form of ${\cal V}_6$ is given by 
\begin{eqnarray}
\label{eq:staggered}
-Ha_0\partial_x\phi/\sqrt{2\pi}-h b_0\cos(\sqrt{2\pi}\theta). 
\end{eqnarray}
The term $\partial_x\phi$, inducing a finite $M$, 
can be absorbed into the free boson part ${\cal H}_0^{\rm eff}$, 
and then the effective Hamiltonian is also a SG model~\cite{O-A}. 
Therefore, we can again apply the form-factor method to calculate 
$I(\omega)$. The soliton mass is given by Eq.~(\ref{eq:solitonmass}) 
with replacing $(K,d)$ by $(1/K,hb_0)$, and the breather masses 
are given by $E_n=2E_s\sin[n\pi/(8K-2)]$ with $n=1,\cdots,[4K-1]$.
Since the value of $K$ increases with increasing $H$, 
the number of breathers $[4K-1]$ is also increased with $H$ 
in the present case. 
From the form-factor method~\cite{K-E}, 
$\partial_x\phi$, $\cos(\sqrt{2\pi}\theta)$, 
and the tilting-bond term $\sin(\sqrt{2\pi}\phi)$ in the Raman operator 
are respectively shown to provide $\delta$-functional peaks 
of odd-th breathers at $\omega=E_{\rm odd}$, 
even-th breathers at $\omega=E_{\rm even}$, 
and soliton (antisoliton) with wavenumber $k=2\pi M/a_0$ 
at $\omega=(E_s^2+k^2v^2)^{1/2}$ in the spectrum $I(\omega)$. 
Namely, in contrast to the case of ${\cal V}_5$, 
all of the elementary particles of the SG model can be observed. 
The $H$ dependence of several peak positions 
are plotted in Fig.~\ref{fig:Gapful} (b). 
Remarkably, level crossings between soliton and breather peaks 
occur~\cite{Umegaki}. 
In addition to these peaks, there exist continuum spectra with 
a smaller weight, although it is difficult to accurately evaluate them. 

In conclusion, we have shown that various weak perturbations ${\cal V}$ 
to the spin-$\frac{1}{2}$ AF Heisenberg Hamiltonian, 
which are expected to determine the quantum dynamics 
in different real 1D antiferromagnets, 
will have distinctive spectral responses in Raman scattering studies of 
1D antiferromagnets. The results summarized in Tables~\ref{tab:table1} 
and \ref{tab:table2} provide 
a means of obtaining useful information about different perturbations 
by comparing our results with future experimental results.

This work is supported by Grant-in-Aids for Scientific 
Research (No. 21244053, No. 21740295, No. 22014016, and No. 23740298) 
from the Ministry of Education, Culture, Sports, 
Science and Technology of Japan, and also 
by Funding Program for World-Leading Innovative R\&D 
on Science and Technology (FIRST Program).


\begin{thebibliography}{99}


\bibitem{Devereaux} See, for a review, T.P. Devereaux, and R. Hackl, 
Rev. Mod. Phys. {\bf 79}, 175 (2007).

\bibitem{Lemmens} See, for a review, P. Lemmens, G. G\"untherodt, and G. Gros, 
Phys. Rep. {\bf 375}, 1 (2003).

\bibitem{Katsura} H. Katsura, M. Sato, T. Furuta, and N. Nagaosa, 
Phys. Rev. Lett. {\bf 103}, 177402 (2009). 

\bibitem{Wang}
See, for a review, K.F. Wang, J.-M. Liu, and Z.F. Ren, Advances in Physics 
{\bf 58}, 321 (2009).





\bibitem{Fleury}
P.A. Fleury and R. Loudon, Phys. Rev. {\bf 166}, 514 (1968).


\bibitem{Moriya}
T. Moriya, J. Phys. Soc. Jpn. {\bf 23}, 490 (1967); 
J. App. Phys. {\bf 39}, 1042 (1968). 




\bibitem{Giamarchi}T. Giamarchi, {\it Quantum Physics in One Dimension} 
(Oxford Univ. Press, New York, 2004). 



\bibitem{Muthukumar}
V.N. Muthukumar, C. Gros, W. Wenzel, R. Valent\'i, 
P. Lemmens, B. Eisener, G. Gu\"ntherodt, M. Weiden, C. Geibel, 
and F. Steglich, Phys. Rev. B {\bf 54}, R9635 (1996). 


\bibitem{Lemmens2}
P. Lemmens, M. Fischer, and G. G\"untherodt, C. Gros, 
P. G. J. van Dongen, M. Weiden, W. Richter, C. Geibel, 
and F. Steglich, 
Phys. Rev. B {\bf 55}, 15076 (1997).

\bibitem{Loosdrecht}
P. H. M. van Loosdrecht, J. Zeman, G. Martinez, G. Dhalenne, 
and A. Revcolevschi, 
Phys. Rev. Lett. {\bf 78}, 487 (1997). 


\bibitem{Ruckamp}
R. R\"uckamp, J. Baier, M. Kriener, M. W. Haverkort, 
T. Lorenz, G. S. Uhrig, L. Jongen, A. M\"oller, G. Meyer, and 
M. Gr\"uninger, Phys. Rev. Lett. {\bf 95}, 097203 (2005).



\bibitem{Schmidt}K.P. Schmidt, C. Knetter, and G.S. Uhrig, Phys. Rev. B 
{\bf 69}, 104417 (2004). 
113 (2000).

\bibitem{Schmidt2}
K.P. Schmidt, C. Knetter, and G.S. Uhrig, 
Europhys. Lett. {\bf 56}, 877 (2001). 


\bibitem{Orignac}E. Orignac, R. Citro, S. Capponi, and D. Poilblanc, 
Phys. Rev. B {\bf 76}, 144422 (2007). 


\bibitem{Lukyanov}S. Lukyanov and A.B. Zamolodchikov, Nucl. Phys. 
B {\bf 493}, 571 (1997).

\bibitem{H-F}T. Hikihara and A. Furusaki, Phys. Rev. B {\bf 58}, 
R583 (1998); Phys. Rev. B {\bf 69}, 064427 (2004). 

\bibitem{T-S}S. Takayoshi and M. Sato, Phys. Rev. B {\bf 82}, 214420 (2010). 





\bibitem{Dender}
D.C. Dender, P.R. Hammar, D.H. Reich, C. Broholm, 
and G. Aeppli, Phys. Rev. Lett. {\bf 79}, 1750 (1997). 

\bibitem{Umegaki}
I. Umegaki, H. Tanaka, T. Ono, H. Uekusa, and H. Nojiri, 
Phys. Rev. B {\bf 79}, 184401 (2009). 

\bibitem{Feyerherm}
R. Feyerherm, S. Abens, D. G\"unther, T. Ishida, 
M. Mei\ss ner, M. Meschke, T. Nogami, and M. Steiner, 
J. Phys.: Condens. Matter {\bf 12}, 8495 (2000).




\bibitem{note1}
We have checked that the peak amplitude of the RSS per one spin 
for the 1D Heisenberg chain ${\cal H}_0$ 
with a small tilting angle (e.g., $\theta_0=\pi/18$) 
is comparable with that for the N\'eel state of the 2D Heisenberg model 
with the same value of the exchange coupling $J$
in the low-temperature region $T/J\ll 1$. 
The latter RSS for the N\'eel state can be evaluated by the standard 
spin-wave theory, and is dominated by two-magnon processes. 
The two-magnon RSSs have been detected in several AF ordered materials 
[See, e.g., M. G. Cottam and D. J. Lockwood, 
{\it Light scattering in Magnetic Solids} (John Wiley $\&$ Sons 1986)]. 
These facts strongly suggest that continuous RSSs can 
also be observed in real quasi-1D quantum magnets. 












\bibitem{E-K}F.H.L. Essler and R.M. Konik, cond-mat/0412421. 

\bibitem{K-E}I. Kuzmenko and F.H.L. Essler, Phys. Rev. B {\bf 79}, 
024402 (2009).





\bibitem{O-A}M. Oshikawa and I. Affleck, Phys. Rev. Lett. {\bf 79}, 
2883 (1997). 



\end{thebibliography}
\end{document}